\documentclass[]{aa} 

\usepackage{url}
\usepackage{graphicx}
\usepackage[varg]{txfonts}
\usepackage{hyperref} 
\hypersetup{breaklinks=true,colorlinks=true,linkcolor=blue,citecolor=blue,urlcolor=blue}
\usepackage{natbib,twoopt}
\bibpunct{(}{)}{;}{a}{}{,}
\usepackage{color}
\usepackage{pdflscape}
\usepackage{todonotes}

\graphicspath{{figures/}}

\newcommand{\kms}{km\,s$^{-1}$}

\newcommand{\califa}{\texttt{CALIFA}}            
\newcommand{\sauron}{\texttt{SAURON}}      
\newcommand{\atlas}{\texttt{ATLAS$^\mathrm{3D}$}} 
\newcommand{\manga}{\texttt{MaNGA}}     
\newcommand{\sami}{\texttt{SAMI}}            
\newcommand{\stan}{\texttt{Stan}}            

\begin{document} 


\titlerunning{BAYES-LOSVD: a bayesian framework for non-parametric extraction of the LOSVD}
\authorrunning{Falc{\'o}n-Barroso \& Martig}

\title{BAYES-LOSVD: a bayesian framework for non-parametric extraction of the line-of-sight velocity distribution of galaxies}

\author{J.~Falc\'on-Barroso\inst{\ref{inst1},\ref{inst2}}\thanks{Email: jfalcon@iac.es}
   \and M. Martig\inst{\ref{inst3}}\thanks{Email: M.Martig@ljmu.ac.uk}
} 
\institute{
Instituto de Astrof\'isica de Canarias, V\'ia L\'actea s/n, E-38205 La Laguna, Tenerife, Spain\label{inst1}
\and
Departamento de Astrof\'isica, Universidad de La Laguna, E-38200 La Laguna, Tenerife, Spain\label{inst2}
\and
Astrophysics Research Institute, Liverpool John Moores University, 146 Brownlow Hill, Liverpool L3 5RF, UK\label{inst3}
}

\date{Received October 9, 2020; accepted}

\abstract{We introduce BAYES-LOSVD, a novel implementation of the non-parametric extraction of line-of-sight velocity distributions (LOSVDs) in galaxies. We employ bayesian inference to obtain robust LOSVDs and associated uncertainties. Our method relies on principal component analysis to reduce the dimensionality of the base of templates required for the extraction and thus increase the performance of the code. In addition, we implement several options to regularise the output solutions. Our tests, conducted on mock spectra, confirm the ability of our approach to model a wide range of LOSVD shapes, overcoming limitations of the most widely used parametric methods (e.g. Gauss-Hermite expansion). We present examples of LOSVD extractions for real galaxies with known peculiar LOSVD shapes, i.e. \object{NGC\,4371}, \object{IC\,0719} and \object{NGC\,4550}, using MUSE and \sauron\ integral-field unit (IFU) data. Our implementation can also handle data from other popular IFU surveys (e.g. \atlas, \califa, \manga, \sami). Details of the code and relevant documentation are freely available to the community in the dedicated repositories.}

         
\maketitle 


\section{Introduction}
\label{sec:intro}

Galaxies are made of stars that move on orbits with different degrees of coherence around their nuclei. Each orbital group contains detailed information about the assembly history of each component and thus retains some memory of the different accretion events suffered by the galaxy over time. This information is encoded in their line-of-sight velocity distribution (LOSVDs), and thus by extracting this property we have access to vital clues to unravel the formation and evolution of galaxies we see today. The analysis of the LOSVD can be done directly in the Milky Way and in the Local Universe by tracing the motions of stars with common stellar populations along a given line-of-sight \citep[e.g.][]{norris86, tolstoy04, deason11, kunder12, ness13, debattista15, zoccali17, du20}.  This is a much more difficult task in nearby galaxies beyond the Local Group, where stars are unresolved and thus the LOSVD at a given position in a galaxy represents multiple populations along that line-of-sight.

The extraction of the LOSVD of galaxies has been an active field of research for many decades. Both parametric and non-parametric approaches have been developed over the years to address this issue. Interestingly original implementations prioritised non-parametric over parametric recoveries, something that has radically changed in the last few decades. Most popular approaches include: {\it Fourier correlation quotient} (FCQ) \citep[e.g.][]{simkin74, sargent77, fi88, bender90}. {\it Cross-correlation (XC or CCF)} \citep[e.g.][]{td79, statler95}. {\it Maximum penalised likelihood} (MPL) \citep[e.g.][]{sw94, merritt97, pinkney03}. {\it Direct fitting in pixel space} \citep[e.g.][]{rw92, km93, vf93, geb00, kelson00, ppxf, ocvirk06, chilingarian07}. 

The extraction of the LOSVD is a degenerate problem, as there are an indefinite number of combinations of stellar populations and LOSVD shapes that can explain a particular spectroscopic observation. Breaking the degeneracies implies a perfect knowledge of the underlying stellar populations that contribute to a particular line-of-sight. In the last few decades, this issue has been mitigated with the advent of a large number of intermediate-resolution stellar libraries and stellar population models \citep[see e.g.][]{bc03, valdes04, coelho05, pat06, prugniel07, vazdekis10, gonneau20, maraston20}, that have helped to greatly reduce the so-called effect of {\it template mismatch} \citep[e.g.][]{fb03}. Another important aspect is the uncertainty of the recovered LOSVD. The methods cited above handle this in different ways. Most of them deal with it with the aid of Monte Carlo simulations by perturbing the input spectrum several times with some known observed uncertainty (but see \citealt{debruyne03} for a more realistic approach). While in recent years the use of parametric forms of the LOSVD has prevailed in the literature \citep[e.g.][]{emsellem11, fb17, vandesande17}, it is becoming increasingly clear that non-parametric approaches are needed to describe the complexity in the LOSVD shapes observed in real data \citep[e.g.][]{jore96, kuijken96, halliday01, gb06, katkov11, coccato13, fabricius14, pizzella18} and numerical simulations \citep[e.g.][]{jesseit07, martig14, schulze17}.

Bayesian inference methods \citep[e.g.][]{nuts} offer a natural way of treating both: (1) the uncertainties in the fitting process, with the advantage that they allow the inclusion of our knowledge of the problem during the fitting process via priors on the input parameters, and (2) the true complex, non-parametric nature of the LOSVDs. \citet{sw94} (hereafter SW94) already thought about this particular way of framing the problem, but limitations in computer performance did not allow them to perform a fully general optimisation of both the LOSVDs and templates. In this paper, we revise the SW94  approach and use the latest developments on Bayesian inference and dimensionality reduction techniques to develop a Python\footnote{\url{https://www.python.org/}} implementation that can efficiently handle simultaneously template optimisation and robust LOSVD extraction. 

The paper is organised as follows. We describe the problem of LOSVD extraction, the techniques for dimensionality reduction of the templates, and LOSVD regularisation in section~\ref{sec:method}. We present our tests on mock spectra in \S\ref{sec:tests}, and apply our extraction methods to real data in \S\ref{sec:data}. We provide all the necessary technical details of our python implementation in section~\ref{sec:implementation}, and provide a summary of the paper and outlook for future applications of this methodology in \S\ref{sec:conclusions}.

\section{The LOSVD extraction} 
\label{sec:method}

A LOSVD represents the distribution of the number of stars as a function of velocity along a particular line-of-sight in a galaxy. In spectroscopic data of nearby galaxies, the LOSVD is the broadening function to be applied to the spectrum of the underlying stellar populations. In essence, the extraction of the LOSVD is a deconvolution problem. 

\subsection{The equation}

In its simplest form, the equation that describes the model to be fit to the data can be expressed in mathematical terms as:
\begin{equation}
    G_{model}\,(\lambda) = \sum_{k=1}^{K} [w_k \cdot T_k(\lambda)] \star B + C(n),
\label{eq:1}
\end{equation}
\noindent where $w_k$ are the weights for each stellar population template ($T_k$), $B$ is the broadening function (i.e. the LOSVD), the $\star$ operator a convolution, and $C(n)$ an additive polynomial of order $n$. The polynomial term is convenient to reduce the impact of template mismatch (which can happen even with the most complete template libraries), and other imperfections during data reduction (e.g. non-perfect sky subtraction and/or scattered light). The equation can become more complicated if the effects of dust attenuation and/or calibration issues between data and templates are to be taken into account (see equation~11 in \citealt{cap17} for a more complete example). The implementation presented in this paper uses the prescription shown in Eq.~\ref{eq:1}, as we have checked that it works for a wide range of datasets, but it can be easily extended to include other correction terms if required. 

The optimisation procedure involves the minimisation of the residuals:
\begin{equation}
   r_{p} = \frac{G_{data}(x_p) - G_{model}(x_p)}{\Delta G_{data}(x_p)},
\label{eq:2}
\end{equation}
\noindent where $x_p$ is the value of the data or model at a given pixel, $G_{data}$ is the observed spectrum, $\Delta G_{data}$ are the observed uncertainties, and $G_{model}$ is the model presented in eq.~\ref{eq:1}. The most commonly used method for the minimisation of eq.~\ref{eq:2} is Least Squares, as there are plenty of computer implementations \citep[e.g.][]{lawson74, more80, jones01} to perform the fits very efficiently.

Inspired by the work of SW94, we have opted for revisiting their bayesian non-parametric approach for the LOSVD extraction. There were three main aspects of SW94 work that were not possible to explore given the computer capabilities and/or mathematical methods available at the time: (1) the Markov Chain Monte Carlo (MCMC) sampling strategy, (2) template optimisation, (3) different forms of regularisation for the LOSVD. We describe the new improvements in the following sections.

\begin{figure*}
   \centering
   \includegraphics[width=\linewidth]{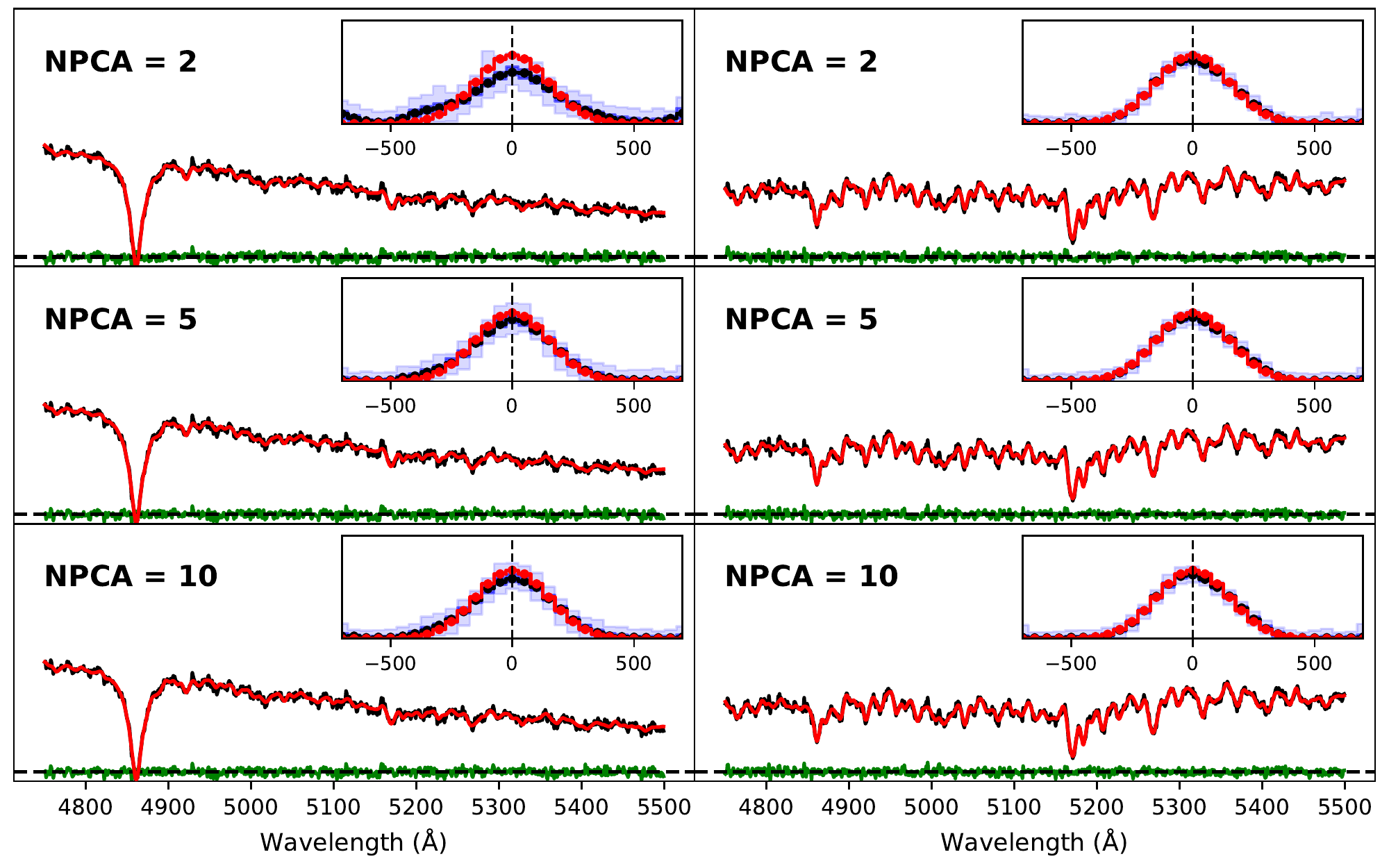}
   \caption{Comparison of the  quality of spectral fitting and LOSVD recovery for different numbers of PCA components. Each column presents input spectra with different stellar populations (young on the left, and intermediate/old populations on the right). The input spectra have a S/N of 50 per pixel. All panels are plotted on the same scale. Black line on main panels are the input test data, while the red line shows the best fitting model. Residuals are indicated in green. The spectral fits are carried out with 2, 5 and 10 PCA templates (from top to bottom as indicated). In the insets, the input LOSVD is a Gaussian centred at zero and a velocity dispersion of 150 \kms\ (indicated in red). Units of the abscissae in the insets are \kms. The recovered median values of the LOSVDs are indicated with a thick black line. 16\%$-$84\% and 1\%$-$99\% confidence limits at each point are indicated in dark and light blue, respectively. An order 2 auto-regressive prior was used to perform the fitting (see \S\ref{sec:priors} for details).}
   \label{fig:pca}
\end{figure*}

\subsection{Markov Chain Monte Carlo sampling}

There are multiple possible strategies for the sampling of parameter space in bayesian inference frameworks. Classical approaches include Metropolis-Hastings \citep{metropolis, hastings}, or Gibbs \citep{gibbs} samplers. With more complex models, the field has experienced a spur of new methods to efficiently probe large parameter spaces: Nested sampling \citep[e.g.][]{buchner14}, Hamiltonian Monte Carlo \citep[e.g.][]{hmc}, or Stein Variational Gradient Descent \citep[e.g.][]{svgd} to cite a few. We refer the interested reader to Chi Feng's Github webpage\footnote{\url{http://chi-feng.github.io/mcmc-demo/}} for a demo on the performance of different samplers.

The SW94 approach relied on the Metropolis algorithm for the exploration of parameter space. Our implementation is based on the No-U-Turn-Sampler (NUTS) introduced by \citet{nuts}, a much more effective sampler. This is part of the \stan\footnote{\url{https://mc-stan.org/}} package \citep{carpenter17}, our software of choice for the MCMC sampling. \stan\ is a probabilistic programming language for statistical modelling and data analysis used in many fields, from physics or engineering to social sciences. We refer the interested reader to \citet{stenning16,asensio17,parviainen18,lamperti19,dullo20} for a few examples of \stan\ applications in astronomy.

Besides the sampling scheme, one of the main advantages of our implementation in \stan\ is the use of the {\it simplex} (i.e. a vector of positive values whose sum is equal to one) to describe the LOSVD. This type of parametrisation provides, naturally, physical and normalisation constraints, and allows for a very efficient exploration of parameter space during minimisation \citep{betancourt12}.\looseness-2

\subsection{Template optimisation}
\label{sec:templates}

A crucial element in the recovery of the LOSVD is the basis of stellar templates used to fit the observed spectra. Ideally, such basis should contain stellar spectra covering the widest range possible of stellar parameters (i.e. T$_\mathrm{eff}$, [Fe/H], log(g), and stellar abundances) or stellar population models with a good sampling of, e.g., ages, metallicities, IMF shapes and slopes, and possibly different chemical abundance ratios too. As already mentioned in the introduction, this is now possible with the advent of latest the stellar libraries and models (see \S\ref{sec:intro} for details).

Using the typical set of $\sim$500-1000 templates, the minimisation involves finding the weights ($w_k$) for each of them, and performing the convolution of the LOSVD on $\sim$1000 spectral pixels per fitting iteration. This is a very time consuming task even for the most efficient bayesian samplers. Since reducing the number of pixels to fit may not be an option, we are thus left with the only alternative of decreasing the number of templates to be used in the optimisation process. This will have the advantage of not only reducing the number of parameters (i.e. $w_k$), but also boosting minimisation performance by very large factors (e.g. from days to minutes). 

There are a number of well known techniques for dimensionality reduction that one could use to whittle down the number of templates, e.g., Independent Component Analysis \citep{lu06}, Factor analysis \citep{nolan06}, Non-Negative Matrix factorization \citep{br07}, Diffusion Maps \citep{richards09}, or $K$-means \citep{sanchez13}. Among all alternatives, we favoured the well-known principal component analysis (PCA) method. PCA has been extensively used in astrophysics for problems very much related to spectral fitting \citep[e.g.][]{ronen99, li05, chen12} and has proven to be both robust and very effective in speeding up calculations. In our particular problem the use of PCA has improved performance from several days to a few minutes (i.e. a 500$\times$ fold decrease of computing time).

In practice the use of PCA means capturing well over 99\%\ of the variance present in a template library of $\sim$1000 spectra with less than 10 PCA components. Figure~\ref{fig:pca} shows the effect of using different number of PCA components during the fit for two mock spectra\footnote{For reference, for this particular case, the continuum-to-line ratio (defined as 100\,$\times\,\sigma_{residuals}/\sigma_{spectrum}$) is 13\% and 23\% for the young and old populations respectively, almost independent of the number of PCA components used in the fitting procedure.} with signal-to-noise ratio per pixel (hereafter S/N) of 50. It is remarkable how a number as low as 2 PCA components can already reproduce with great accuracy the observed spectrum. This is, in part, due to the use of additive polynomial that helps to reduce potential template mismatch issues. We have checked the recovery for different types of stellar populations (as shown in the two columns) and the extracted LOSVD is less accurate for young stellar populations with a low number of PCA components. This not unexpected given that spectra of younger populations show greater variation, which is difficult to capture with just 2 PCA components. We noticed that the closer the input spectrum is to the average spectra of the templates (i.e. intermediate stellar populations in our particular case), the better the recovery is for low number of PCA components. In general, our tests suggest that 5  PCA components suffice to recover the input LOSVD with great accuracy, with a clear improvement as S/N increases.\looseness-2

The optimal selection of the number of PCA components is not a well defined quantity though. Choosing a number of PCA components that would explain a certain variance of the spectra (e.g. compatible with the noise of the spectrum) seems a reasonable approach. However, in practice, for real data, this means selecting a different number of components for each spectrum of the dataset, which is very impractical. In our experience, for typical S/N ratios between 50 and 100, it is sufficient to choose a number between 5 and 10 PCA components\footnote{This is using the MILES models \citep{vazdekis10}}. In our implementation, this number is a free parameter that the user has to decide on before runtime.\looseness-2

\subsection{LOSVD regularisation}
\label{sec:priors}

The extraction of the LOSVD is a degenerate problem with a large number of LOSVD shapes and combinations of templates that can reproduce the observed spectrum with great accuracy. Nevertheless the range of allowed LOSVD solutions can be constrained for reasonably high S/N data. 

There are different ways to impose some level of regularisation onto the output LOSVD. In our bayesian approach they are expressed in the form of priors over the LOSVD. We have explored four different types of priors:
\begin{enumerate}
   \item No regularisation 
   \begin{equation}
      \mathrm{LOSVD}_i\sim\mathcal{N}(0,\sigma^2),
      \label{eq:3}
   \end{equation}    
   \noindent where $\mathrm{LOSVD}_i$ is the i-th velocity element of the LOSVD, and $\mathcal{N}(0,\sigma^2)$ represents a normal distribution with mean zero and variance $\sigma^2$. This is a fairly uninformative prior that assumes the same level of uncertainty on all LOSVD elements.\newline

   \item Random-Walk prior
   \begin{equation}
      \mathrm{LOSVD}_i\sim\mathcal{N}(\mathrm{LOSVD}_{i-1},\sigma^2),
      \label{eq:4}
   \end{equation}
   \noindent where the i-th element of the LOSVD is linked to the previous one. This type of prior was the one proposed by SW94. We set the prior for the first element to: $\mathrm{LOSVD}_0\sim\mathcal{N}(0,\sigma^2)$.\newline    

   \item Auto-regressive prior
   \begin{equation}
      \mathrm{LOSVD}_i\sim\mathcal{N}\left(\alpha + \sum_{k=1}^{N}\beta_k\times\mathrm{LOSVD}_{i-k},\sigma^2\right).
      \label{eq:5}
   \end{equation}
   \noindent This is a more general form of prior that becomes equal to the Random-Walk for $\alpha=0$ and $\beta=1$ for $k=1$. Stronger regularisation is obtained by increasing the order $k$, which links more consecutive LOSVD elements. We impose the same prior used for the Random-Walk approach to the first element of the LOSVD, and use weakly informative normal priors for $\alpha$ and $\beta$.\newline    
   
   \item Penalised B-splines\newline
   
   \noindent B-splines are a special type of piecewise polynomials controlled by {\it knots} that are often used for interpolation \citep[e.g.][]{press03}. One of the major difficulties on the definition of B-splines is the choice of the number of {\it knots} to define the polynomial function. In our {\tt Stan} implementation, we followed the example provided by Milad Kharratzadeh\footnote{\url{https://mc-stan.org/users/documentation/case-studies/splines_in_stan.html}}, and rather than establishing a relation between different elements of the LOSVD, we apply Random-Walk priors on the B-splines coefficients such that:
   \begin{equation}
      a_1\sim\mathcal{N}(0,1),~~~~a_i\sim\mathcal{N}(a_{i-1},\tau),~~~~\tau\sim\mathcal{N}(0,1).
      \label{eq:6}
   \end{equation}
   \noindent This approach promotes smoothness in the resulting LOSVD, preventing  excessive wiggling in the solution. Another important parameter that controls the level of flexibility of the B-splines is the B-spline order $k$. 
\end{enumerate}

\begin{figure}
   \centering
   \includegraphics[width=\linewidth]{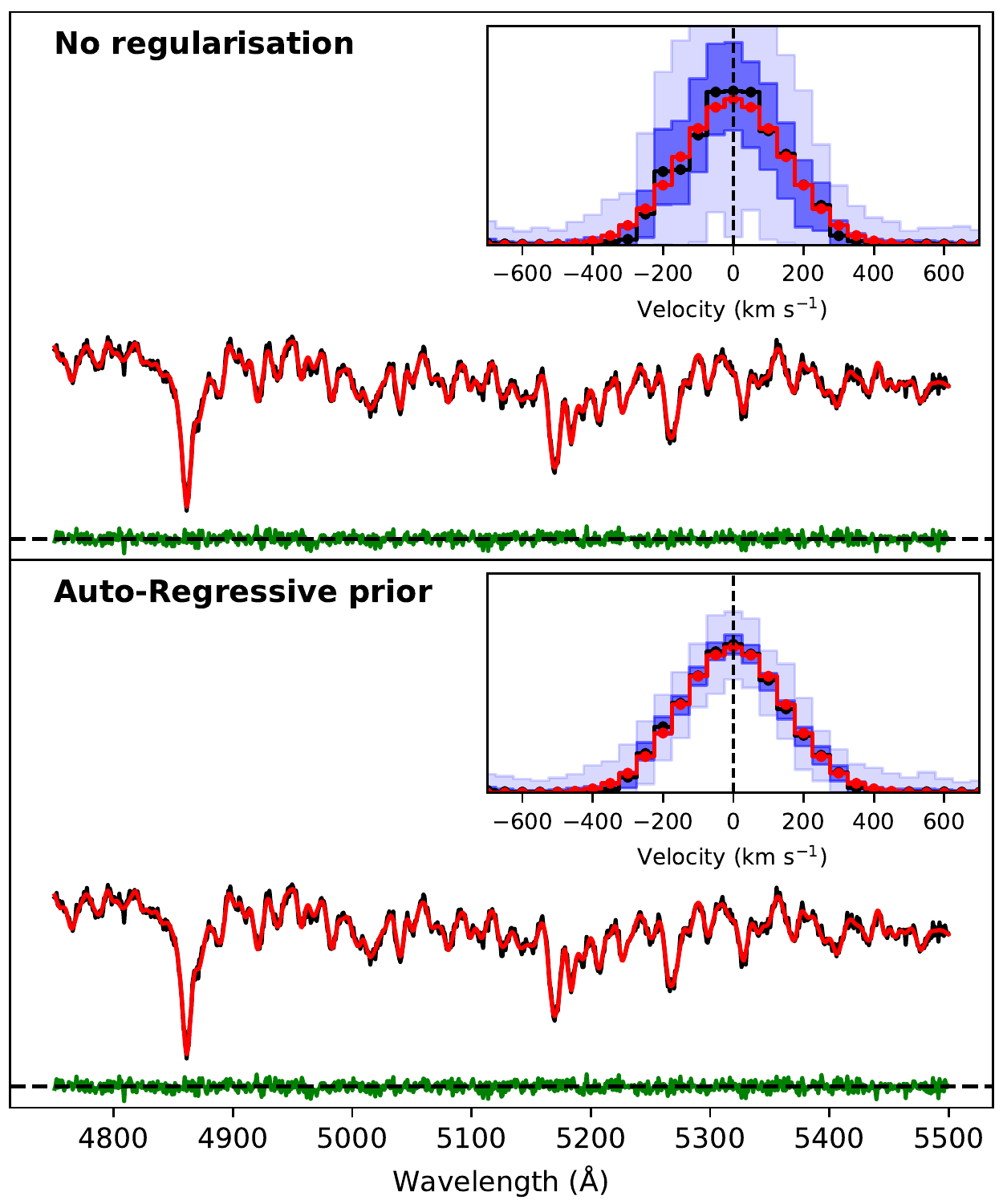}
   \caption{LOSVD recovery for no regularisation and an auto-regressive (order 2) prior. Colors as in Fig.~\ref{fig:pca}}
   \label{fig:priors}
\end{figure}

An interesting feature of our procedure is that we do not impose any predefined value to $\sigma^2$ or $\tau$ during the fitting process. In fact, they are considered nuisance parameters and we let the quality of the data establish their optimal distributions. Far from being unconstrained, it turns out that both parameters are well behaved and display fairly tight distributions. 

We illustrate the effect of the choice of prior for the LOSVD and its uncertainties in Figure~\ref{fig:priors}. We have created a test spectrum for this purpose with a Gaussian LOSVD and a S/N=100 per spectral pixel. The figure shows the difference between no regularisation and an  auto-regressive prior of order 2. Both approaches deliver an indistinguishable fit to the input spectrum, and capture the Gaussian nature of the input LOSVD. However, the level of uncertainty displayed by the case without regularisation is far larger than that of the auto-regressive prior. These results are very much in agreement with \citet{sw94} findings. As we will show in Section~\ref{sec:tests}, this difference persists even at the highest S/N levels and is related to the number of degrees of freedom of one method versus the other.

\begin{figure}
   \centering
   \includegraphics[width=\linewidth]{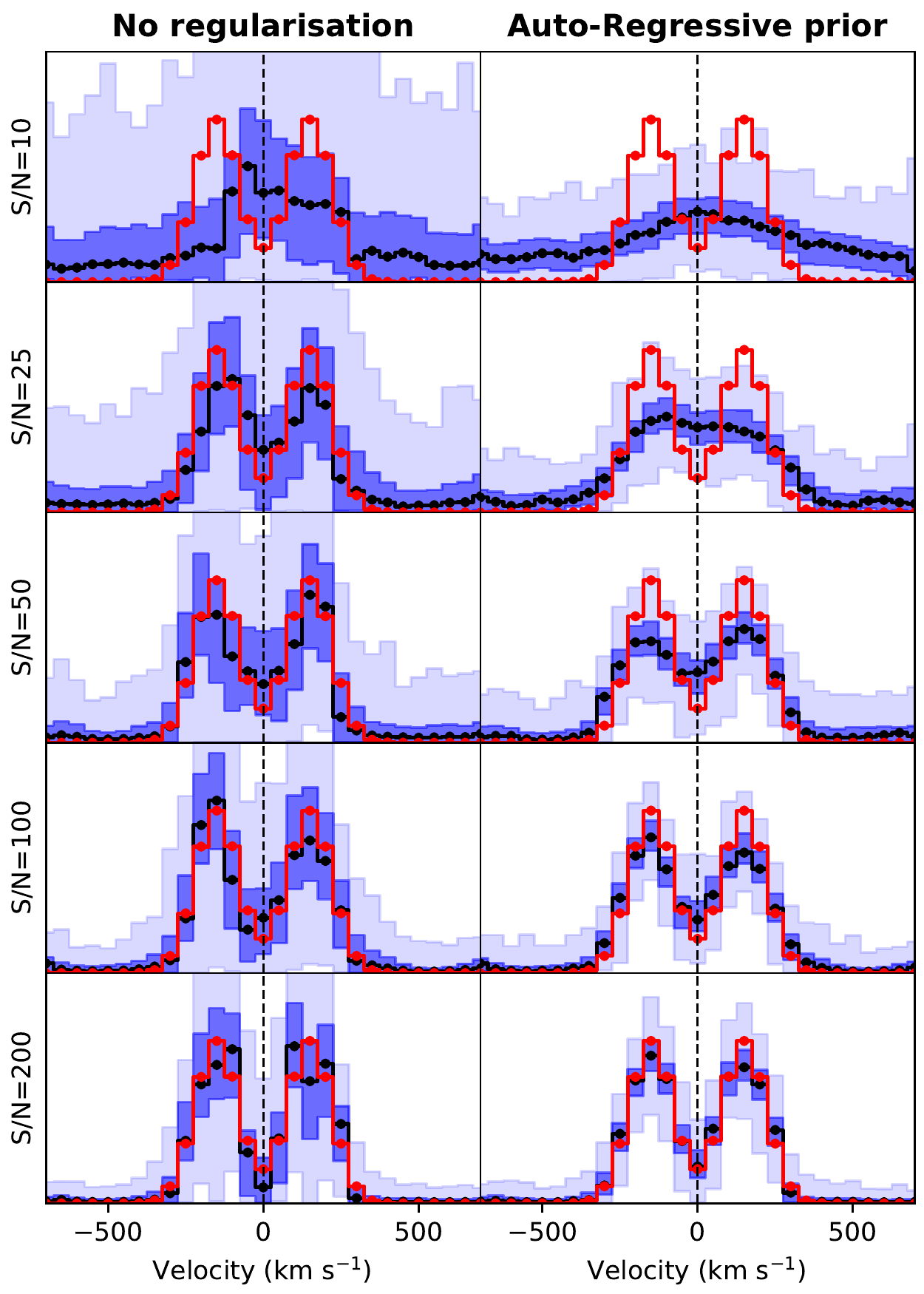}
   \caption{LOSVD recovery for different S/N ratios and types of regularisation. We use a double Gaussian LOSVD profile as an example and it is indicated in red. Left column displays solutions for extraction with no regularisation, while right column shows results for an order 2 auto-regressive prior. Each row represents two cases for a different S/N per pixel as indicated. All panels are plotted on the same scale. Colors as in Fig.~\ref{fig:priors}.\looseness-2}
   \label{fig:snr_recovery}
\end{figure}

\subsection{Likelihood}
\label{sec:likelihood}

Besides the priors, for the minimisation process, we need to define the form of the likelihood of our data given a model.  In our problem we assume that our spectroscopic observations can be explained by a normal distribution such that:

\begin{equation}
   G_{data}\,(\lambda)\sim\mathcal{N}(G_{model}\,(\lambda),~\sigma_{G_{data}}^2),
\label{eq:7}
\end{equation}

\noindent where $G_{data}\,(\lambda)$ is the observed input spectrum, $G_{model}\,(\lambda)$ is our model, and $\sigma_{G_{data}}^2$ is the variance of the observed spectrum. 

Equation~\ref{eq:1} is adequate to define our model when a full template library is to be used during the minimisation process. In our case, however, the use of PCA components transforms that equation to: 

\begin{equation}
   G_{model}\,(\lambda) = \left[\widetilde{T} + \sum_{k=1}^{K} w_k \cdot PCA_k(\lambda)\right] \star B + C(n),
\label{eq:8}
\end{equation}

\noindent where now $w_k$ are the weights for each PCA component (PCA$_k$), $\widetilde{T}$ is the mean template of the input library, $B$ is the broadening function (i.e. the LOSVD), the $\star$ operator a convolution, and $C(n)$ an additive polynomial of order $n$. Our implementation in \stan\ supports both forms of equations, but it is significantly more efficient with Eq.~\ref{eq:8}.

\section{Tests on simulated data}
\label{sec:tests}

We have checked the performance of our implementation under different circumstances by creating mock spectra for a wide range of S/N ratios and input LOSVD shapes. While some of the LOSVDs were arbitrarily defined by combining Gaussian functions, others come from numerical simulations and are thus more realistic. We choose to illustrate here a case with an input spectrum with an intermediate-age stellar population, but results are consistent when other populations were used. Our fits were performed using 5 PCA components.\looseness-2

\begin{figure*}
   \centering
   \includegraphics[width=\linewidth]{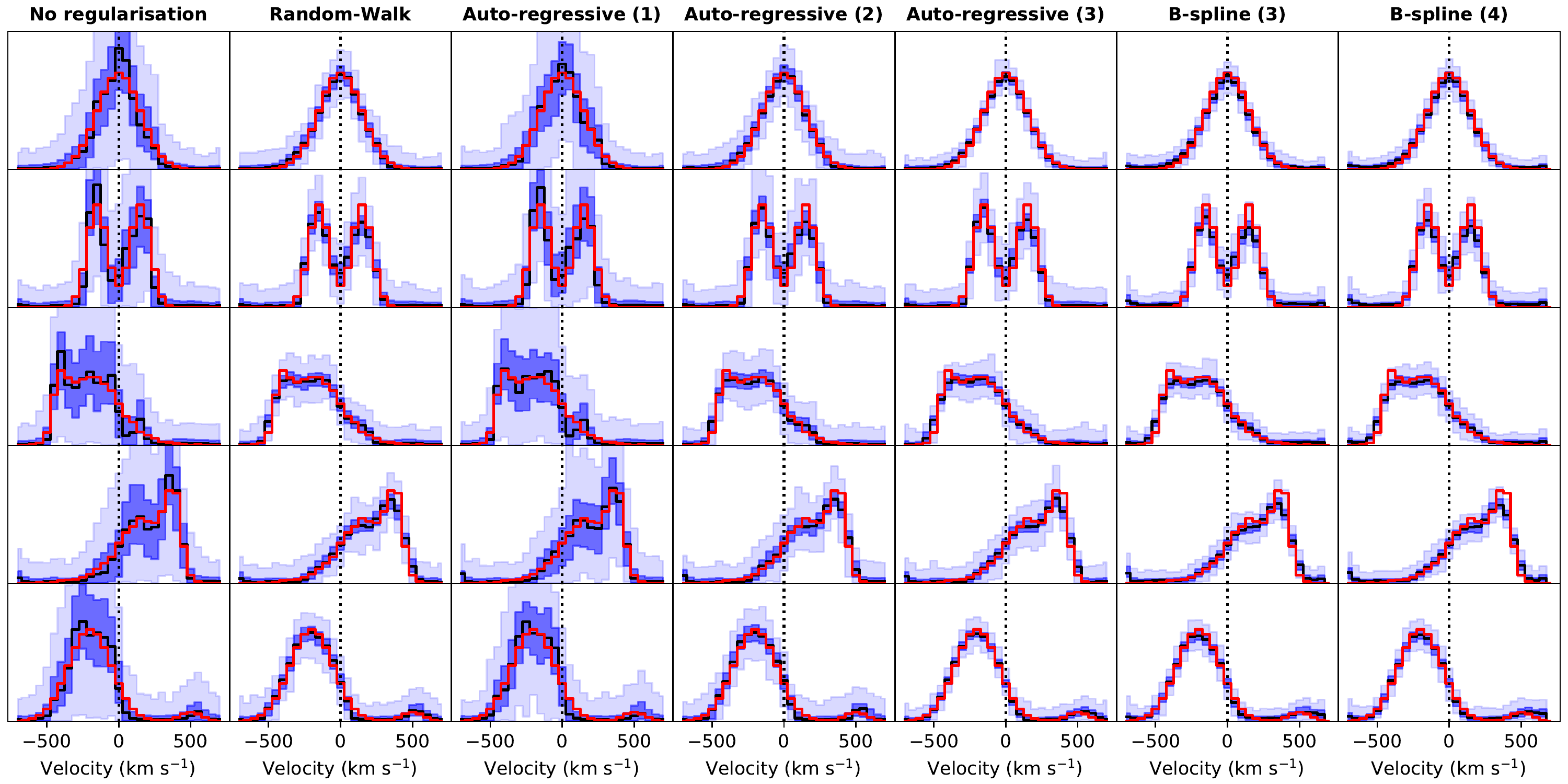}
   \caption{LOSVD recovery for different input LOSVDs and types of regularisation. Colors as in Fig.~\ref{fig:priors}. Each row represents a particular LOSVD shape for different types of regularisation. These are solutions for input spectra with S/N=50 per pixel. All panels are plotted on the same scale.}
   \label{fig:losvd_types}
\end{figure*}

Figure~\ref{fig:snr_recovery} shows an example of the recovery of an extreme case of LOSVD shape (i.e. a double Gaussian profile) for a range of S/N ratios and two prior assumptions. The effect of regularisation is evident from the lowest to the highest S/N ratios. With no regularisation there is a large range of possible solutions, as displayed by the confidence intervals, even at S/N=200. It is interesting to note that the level of uncertainty decreases drastically with S/N when no regularisation is applied, while the improvement is not so large when priors are used. Regularised solutions come with the price of introducing some bias (see \S\ref{sec:data} for an example). This is most acute for the lowest S/N ratios, where the regularised solution fails to capture the double peaked nature of the input LOSVD. It is clear though that at S/N=10, even non-regularised solutions cannot reproduce the input LOSVD. It seems that at S/N of 25 the non-regularised solution clearly reveals already the double peak feature, while the regularised one does not. This is an important result, as it shows that non-regularised solutions are more accurate at low S/N regimes, at the expense of larger uncertainties.  

In Figure~\ref{fig:losvd_types} we present how the different types of regularisation methods listed in \S\ref{sec:priors} influence the recovery of the range of input LOSVDs we prepared for these tests. The S/N ratio of the input data is 50 per spectral pixel. The first thing to notice is that the overall level of accuracy in the LOSVD recovery is rather good (i.e. the input LOSVD is contained within the confidence levels) for all LOSVD shapes. Nevertheless, as expected from previous figures, the use of different priors result in different confidence intervals. An interesting feature is the difference in the confidence intervals between Random Walk and Auto-Regressive (order 1) priors. This is entirely driven by the reduced number of degrees of freedom of the former over the latter (see \S\ref{sec:priors}). Perhaps one of the most important messages from the figure is that subtle details can be extracted at S/N=50. This is particularly striking for the case displayed in the bottom row. This LOSVD is made of two Gaussians: a prominent one centred at $-200$\,\kms, and a faint component located at 500\,\kms. The goal of this test was to check the ability of the code to recover the presence of potential small satellites being accreted onto a galaxy. This appears to have been achieved for any choice of regularisation, with slightly better recovery of the faint component with the Random-Walk and Auto-Regressive priors. These results are quite encouraging, as it means that there is not need for very high S/N ratios to detect this kind of features. We provide versions of Fig.~\ref{fig:losvd_types} for different S/N rations in Appendix~\ref{app:losvd_types}.

\section{The implementation with pyStan}
\label{sec:implementation}

There are many alternatives one could choose from to implement all the ideas proposed in section~\ref{sec:method}. Specifically in Python some of the most popular packages include {\it emcee}\footnote{\url{https://emcee.readthedocs.io/}} \citep{emcee} and pyMC3\footnote{\url{https://docs.pymc.io/}} \citep{pymc3}, but see Gabriel Perren's webpage\footnote{\url{https://gabriel-p.github.io/pythonMCMC/}} for a comprehensive list of options. We picked pyStan\footnote{\url{https://pystan.readthedocs.io}} as our package of choice to develop BAYES-LOSVD, as it offers a convenient python interface to \stan.

BAYES-LOSVD is built in a modular fashion to make it very simple for the end user to extend its capabilities. This includes the addition of: (1) read-in routines for data of a new instrument, (2) new \stan\ models with different kinds of regularisation, (3) new template libraries. Our current implementation supports datacubes from the MUSE-WFM, \sauron, \atlas, \califa, \manga, and \sami\ surveys, besides the possibility to read standard 2D FITS format files with spectra along rows. The linearly-sampled input spectra is preprocessed before execution allowing the user to pick the level of Voronoi binning (using \citealt{voronoi} implementation), and velocity range and sampling of the output LOSVD. In this process, the designated templates will be prepared accordingly, with the possibility of switching off the reduction of the template basis with the PCA scheme described in Sect.~\ref{sec:templates}.

On execution, the user can decide whether to analyse the entire set of Voronoi bins or a just a selection of them. Distributed computing is implemented natively, so that multiple spectra can be executed in parallel on multi-CPU machines. On output, by default, only summary statistics are stored on disk. MCMC posterior distributions are described with highest density interval estimators \citep[e.g.][]{kruschke14}, which are more accurate to describe highly skewed distributions (as opposed to the standard percentiles approach). They are stored on disk in HDF5 format\footnote{\url{https://www.h5py.org/}}. Diagnostic plots are also created if requested. Users wanting to delve into the details can chose to save the entire posterior distribution values, which can be then easily analysed using the Arviz\footnote{\url{https://arviz-devs.github.io/arviz/}} package.  

Performance and execution times depend very much on the data, and parameters used for the LOSVD extraction. Thanks to \stan\ though, convergence is usually achieved with a very small number of iterations. In all the tests presented here 3 chains with 500 iterations (i.e. warm-up + sampling) sufficed to obtain well-behaved posterior distributions. A typical spectrum with $\sim$500 pixels (e.g. 4800$-$5500\,\AA\ region) and S/N=50 per pixel, a velocity range of $\pm$\,700\,\kms with a sampling of 50\,\kms, and 5 PCA components would require $\sim$10\,min on a cluster with Intel Xeon E5-2630 (v4) CPUs. 

There is one major bottleneck in our implementation: convolution is performed in direct space given \stan's current inability to handle complex numbers. This has a very strong impact when wide spectral ranges are to be fitted. Based on discussions on the \stan\ forum\footnote{\url{https://discourse.mc-stan.org/}} we are aware that Fast Fourier Transforms will be possible in the not-so-distant future. Another aspect where we could already optimise performance is the likelihood evaluation. The latest version of \stan\ has introduced new features that can speed this process by large factors by distributing its computation over many CPUs. This is unfortunately not available in the pyStan version (v2.19) we are using, but it will be available with the release of pyStan (v3). Therefore there is still room for performance improvements in the midterm. We will be alert and update the code to keep up with any new development.

The code can be accessed via the Github repository \url{https://github.com/jfalconbarroso/BAYES-LOSVD}. Detailed documentation can be found on the same page.

\begin{figure}
   \centering
   \includegraphics[width=\linewidth]{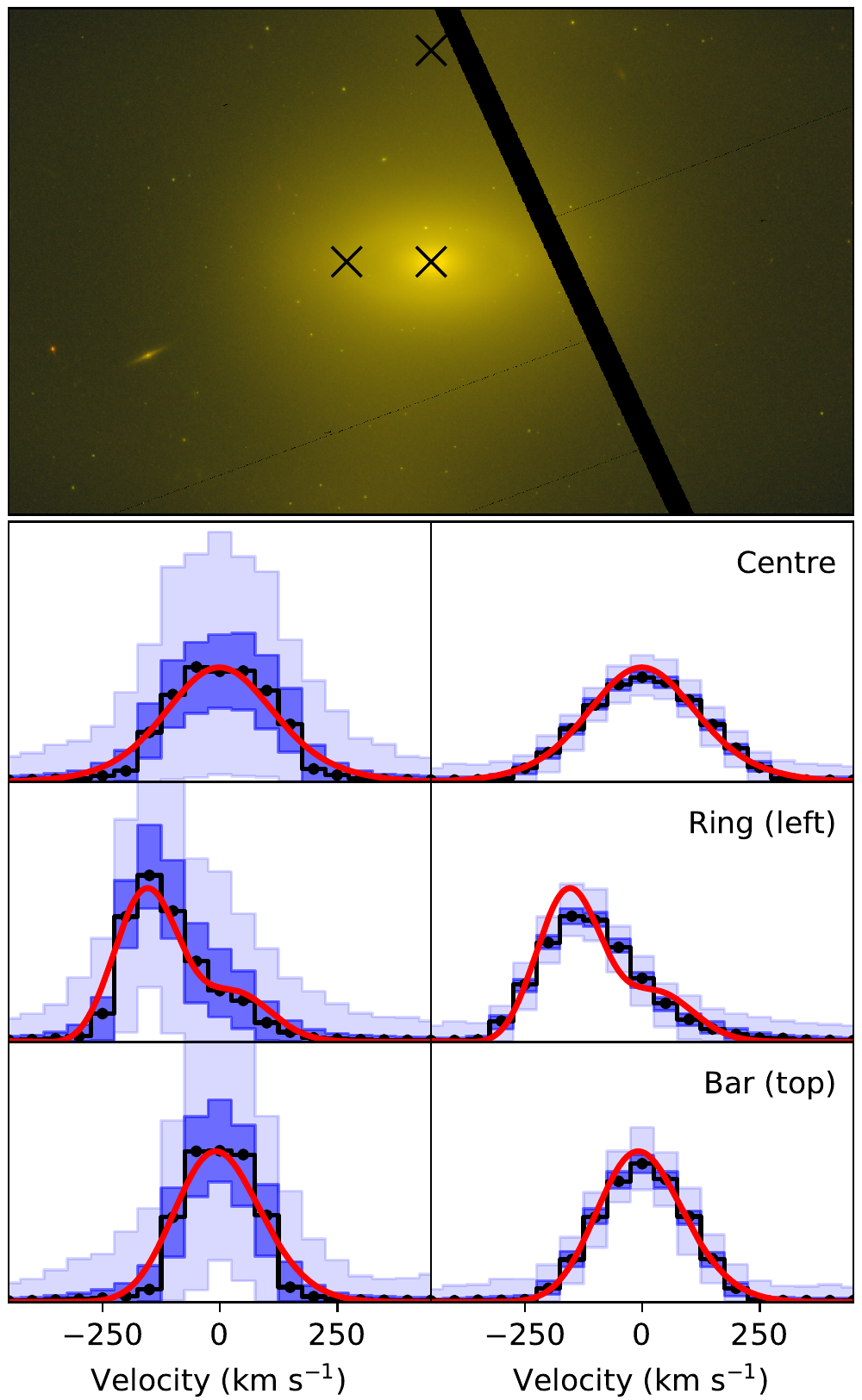}
   \caption{BAYES-LOSVD extraction for NGC\,4371. (Top panel) Hubble Space Telescope colour image based on the F475W and F850LP filters. North is up and East to the left. (Bottom panels) LOSVDs extracted at different locations (as indicated with black crosses on the image). On the left, without regularisation, and on the right with an order 2 auto-regressive prior. Colors as in Fig.~\ref{fig:priors}. Red lines correspond to the best Gauss-Hermite LOSVD extracted with pPXF (see \S\ref{sec:ngc4371} for details).} 
   \label{fig:ngc4371}
\end{figure}

\section{Application to real data}
\label{sec:data}

We now turn the attention to the recovery of LOSVD using real data from different instruments. We show here examples of LOSVDs for three galaxies presenting LOSVDs with varied degrees of complexity. We chose data from MUSE-WFM and \sauron\ IFUs, but note that the code has been benchmarked with data from some of the most popular IFU surveys (e.g. \atlas, \califa, \manga, \sami) too.

\subsection{NGC\,4371}
\label{sec:ngc4371}

The first case we present is NGC\,4371. This galaxy was studied in great detail by \citet{gadotti15} using the MUSE IFU, and was part of the pilot programme for the TIMER survey \citep{gadotti19}. NGC\,4371 is interesting for being a fairly old system (i.e. $\ge$\,7\,Gyr throughout) with evidence for a {\it fossil} nuclear stellar ring void of star formation \citep[e.g.][]{erwin01}. Following \citet{gadotti15} study, we extracted spectra with a 3.0\arcsec aperture in three positions of the galaxy: at the centre, the nuclear stellar ring, and a location along the bar. The S/N of the spectra in the three apertures is well above 100 per pixel. We impose a velocity sampling of 50\,\kms.\looseness-2  

Figure~\ref{fig:ngc4371} shows the results of our analysis. The three LOSVDs are different from each other in different aspects. While the LOSVDs at the centre and bar regions display fairly Gaussian profiles, the ring-dominated region is clearly asymmetric and skewed. The panels on the left hand side show solutions with no regularisation, and the ones on the right used an order 2 auto-regressive prior. In essence, we see the same behaviour observed with the test data in previous sections with non-regularised solutions giving larger confidence intervals than the regularised ones. The two sets of solutions are very much consistent with each other, as also observed in our experiments. These results presented in the figure have been obtained applying our method to spectra in the wavelength range 4800$-$5300\AA. We also computed solutions based on spectra around the Calcium Triplet region (8450$-$8700\AA) and obtained identical solutions (and thus not shown here). This not totally unexpected since there is no evidence for multiple stellar populations in this galaxy. 

In addition, for comparison, we plot the best Gauss-Hermite LOSVD extracted with pPXF \citep{cappellari17}. In this particular case, the agreement between the recovered LOSVDs with our method and pPXF is very good. This is specially true for the almost Gaussian LOSVDs at the centre and bar locations of the galaxy. At the ring, the regularised solution does not match the pPXF result as closely as the non-regularised one, but differences are still within a 1\%$-$99\% percentiles of our non parametric extraction. While for cases like this one, the advantage of the non-parametric approach may not seem evident, it is important to note that the Gauss-Hermite parametrization allows for negative values on the wings of the LOSVD. This situation occurs for h$_3$ and h$_4$ values of 0.1, $-0.1$ respectively, which are not uncommon in the kinematic maps presented in many IFU surveys. Since we constrain the LOSVD during the fit to admit only positive values, our method overcomes this limitation of and provides naturally physically meaningful LOSVDs.  

\begin{figure}
   \centering
   \includegraphics[width=\linewidth]{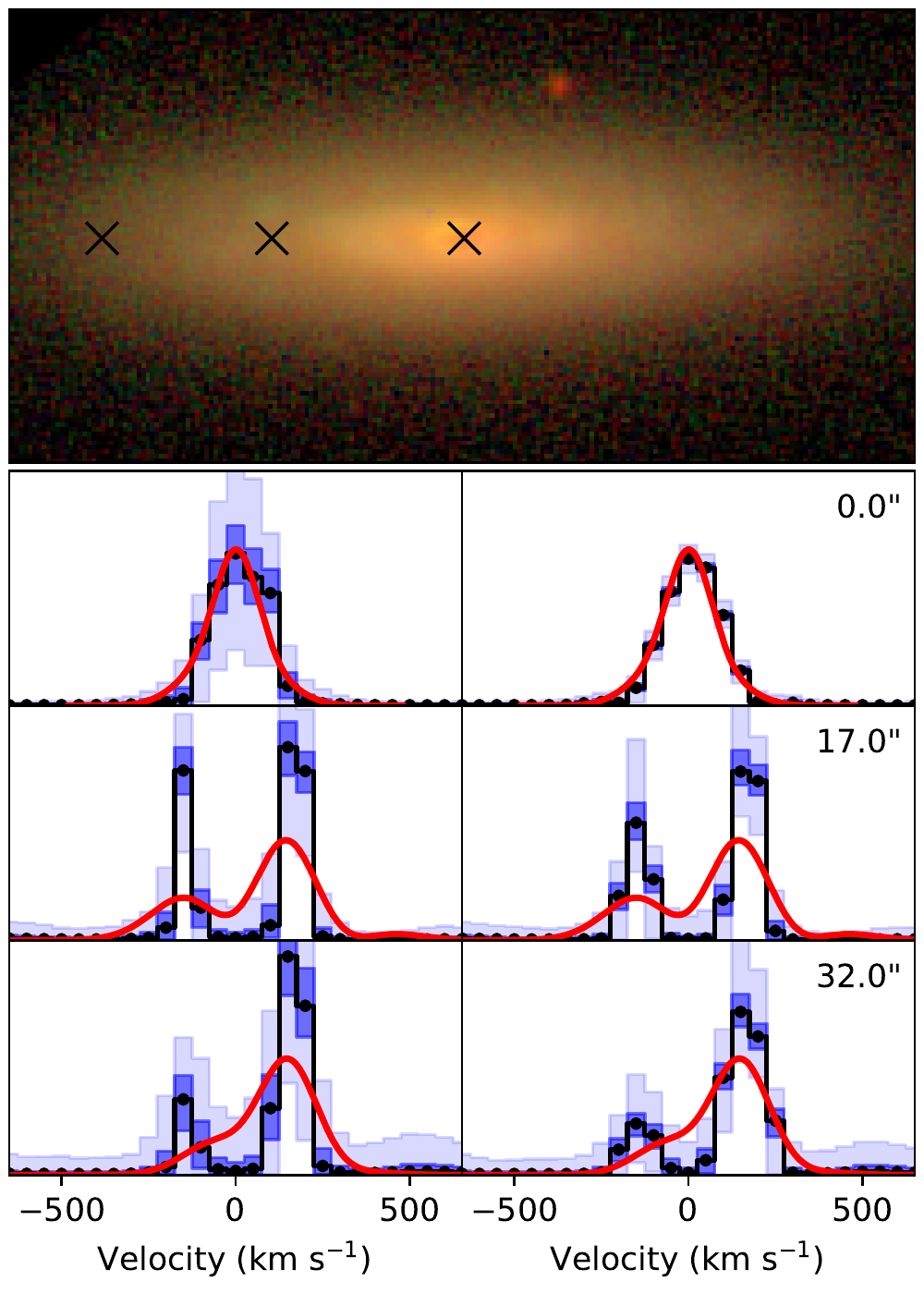}
   \caption{BAYES-LOSVD extraction for IC\,0719. (Top panel) Sloan Digital Sky Survey colour image based on the $g,r,i$ filters. The image has been rotated so that the major axis is parallel to the abscissae. (Bottom panels) LOSVDs extracted at different locations along the major axis of the galaxy (as indicated with black crosses on the image). On the left column, without regularisation, and on the right with an order 2 auto-regressive prior. Colors as in Fig.~\ref{fig:priors}. Red lines correspond to the best Gauss-Hermite LOSVD extracted with pPXF (see \S\ref{sec:ic0719} for details).}
   \label{fig:ic0719}
\end{figure}

\begin{figure*}
    \centering
    \includegraphics[width=\linewidth]{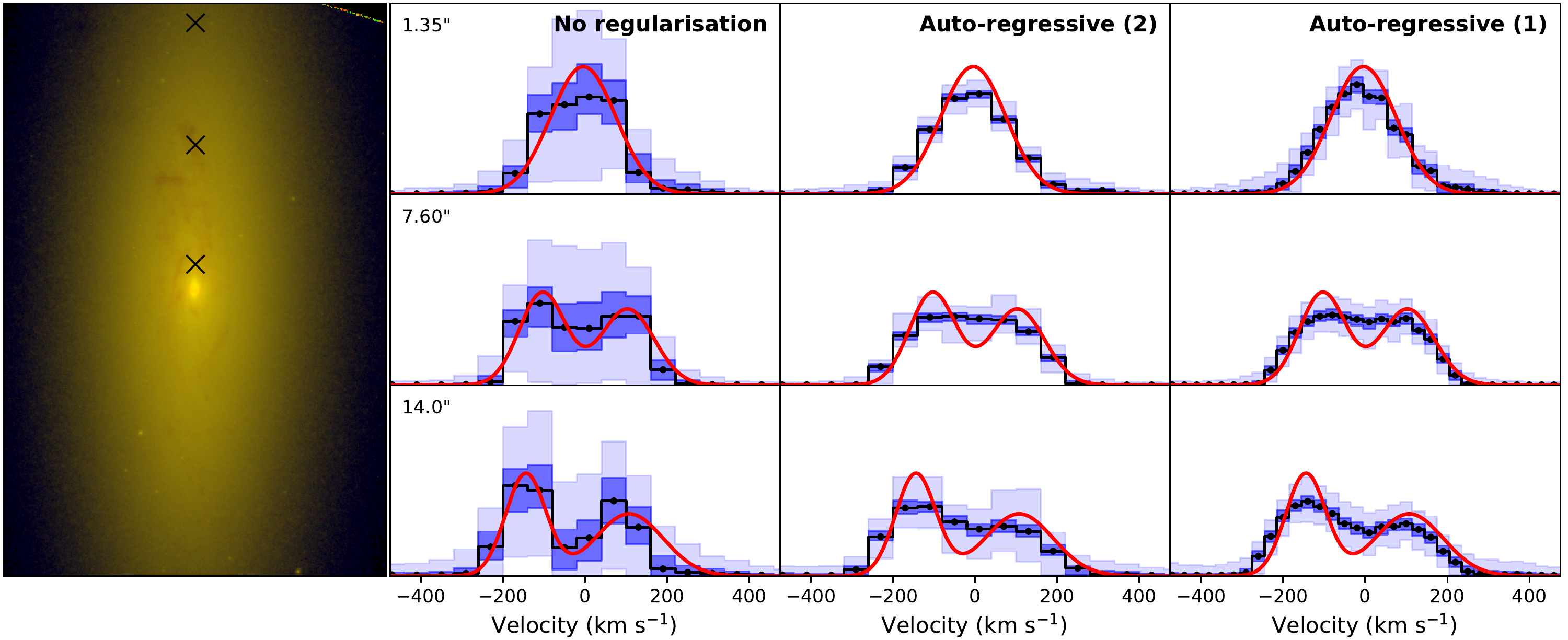}
    \caption{BAYES-LOSVD extraction for NGC\,4550. (Left panel) Hubble Space Telescope colour image based on the F555W and F814W filters. North is up and East to the left. (Other panels) LOSVDs extracted at several locations along the major axis of the galaxy (as marked by the black crosses). Each row corresponds to a different location from the centre of the galaxy (in arcsec) as shown on the first column. Each column uses a different kind of regularisation, as indicated on the top row. Colors as in Fig.~\ref{fig:priors}. Red lines show \citet{rw92} LOSVDs at those locations.}
    \label{fig:ngc4550}
 \end{figure*}
 
\subsection{IC\,0719}
\label{sec:ic0719}

The second case we study is IC\,0719, a spectacular case displaying multiple kinematic components \citep{pizzella18}. The galaxy is made of stars in two counter-rotating large scale discs with distinct stellar populations. In addition it has an ionised-gas component showing the same sense of rotation of the secondary, lower-mass, younger stellar disc. We used MUSE observations around the 4800$-$5300\AA\ region to extract three apertures along the major axis of the galaxy. As for the case of NGC\,4371, the S/N of the spectra is well above 100 per pixel. We impose a velocity sampling of 50\,\kms.

Figure~\ref{fig:ic0719} shows an almost perfect Gaussian LOSVD profile at the centre of the galaxy, while LOSVDs along the major axis display clear double-peaked shapes. This is in perfect agreement to the analysis performed by \citet{pizzella18}. For comparison, we also plot the best fit Gauss-Hermite LOSVD parametrisation obtained with pPXF (in red). Here it becomes obvious that the Gauss-Hermite expansion cannot reproduce well such complicated shapes and have LOSVD wings that have slightly negative values. Although not obvious due to the normalisation, for the aperture at 17\arcsec the pPXF extraction creates a third peak in the LOSVD at velocities $\sim$500\,\kms, where the non-parametric approach goes effectively to zero. It is worth highlighting the level of complexity of the recovered LOSVDs despite the smooth morphological appearance. The same applies to NGC\,4550, as we discuss in Sec.~\ref{sec:ngc4550}, and emphasizes the need for the non-parametric LOSVD extraction in galaxies. This topic is gaining the attention in the literature and non-parametric LOSVDs are now being routinely included in the dynamical modelling of early-type galaxies \citep[e.g.][]{mehrgan19, neureiter20}.\looseness-2 

Another interesting aspect to explore in this galaxy is the non-parametric extraction of the LOSVD in wavelength regions sensitive to different stellar populations. This is actually possible with MUSE data and it will be a matter of analysis in Rubino et al. (in preparation). 

\subsection{NGC\,4550}
\label{sec:ngc4550}

The last case we analysed is NGC\,4550, another classical showcase galaxy with prominent double-peaked LOSVD profiles. We have extracted LOSVDs from \sauron\ spectra \citep{emsellem04}. We performed our calculations in the wavelength range between 4800$-$5300\AA  \ and a S/N=150 per spectral pixel. We analysed the results along the major axis of the galaxy at three of the locations presented in \citet{rw92}. Since their data was not in electronic form, we digitised it using the WebPlotDigitiser\footnote{\url{https://automeris.io/WebPlotDigitizer}} tool \citep{WebPlotDigitizer} and then fitted the LOSVDs with double Gaussian profiles for best reproduction. Figure~\ref{fig:ngc4550} (top panel) shows the F555W/F814W color image from the Hubble Space Telescope. Remaining panels show our recovered LOSVDs at the three positions along the major axis, as indicated. Each row corresponds to a location, while each column uses different priors for the LOSVD extraction: no regularisation, auto-regressive (order 2), auto-regressive (order 1). The red lines show the \citet{rw92} results. 

The first thing to notice is the excellent agreement between our LOSVDs and those of \citet{rw92} when no regularisation is used. The complexity of the LOSVDs increases as we move away from the centre of the galaxy, becoming double-peaked from 7.6\arcsec. The \citet{rw92} results are well within our 16\%$-$84\%\ confidence intervals (dark blue shaded region) despite the different sampling in velocity. 

As opposed to IC\,079, the situation is drastically different when regularisation is applied though. Our auto-regressive (order 2) solutions are not capable of capturing the double-peaked nature of the LOSVDs at larger distances from the centre. We have investigated the source for this discrepancy and concluded that it is related to the intrinsic difference in velocity between the two peaks and the velocity sampling used to extract the LOSVDs. In other words, the level of correlation between velocity bins imposed by this prior is too strong and smooths the solution too much. In order to check this we have extracted the LOSVDs with an auto-regressive (order 1) prior, but sampling the LOSVD in steps of 30\,\kms\ instead of the 60\,\kms\ used in all the other extractions. This is shown in the right-most column. It clearly shows that the two peaks can be recovered with a less stringent prior and finer sampling in velocity. 

Based on these findings, we warn the reader to understand the implications of using regularisation in their analysis. We therefore recommend potential users to perform non-regularised fits to their data, and consider carefully the velocity sampling to be used in the LOSVD extraction.

\section{Summary \& Conclusions}
\label{sec:conclusions}

The advent of very high quality data from many integral-field spectrographs and surveys has opened the possibility of efficiently extracting LOSVDs in galaxies. At the same time, great progress in computer performance, algorithms and mathematical methods, makes it possible now to handle large datasets. Inspired by the work of SW94, we have developed a bayesian inference approach to the LOSVD extraction from spectra. The code improves on SW94 work on 3 main aspects: (1) the Markov Chain Monte Carlo sampling strategy, (2) the possibility of different forms of regularization for the LOSVD, and (3) template optimisation based on PCA components. Our tests on mock data indicate that LOSVD recovery is accurate for spectra with S/N$>$50 with as few as 5 PCA templates. Regularised solutions provide less uncertain LOSVDs, but it is at the expense of biased solutions for low S/N ratios. We have also successfully applied our approach to MUSE and \sauron\ data, displaying many interesting features and warning the users to be careful with regularised solutions in some situations (see \S\ref{sec:ngc4550} for an example). The use of non-regularised solutions should therefore be preferred as it provides non-biased solutions.\looseness-1

On the technical side, our implementation is very versatile and allows the possibility of extending its capabilities in different fronts (i.e. inclusion of read-in routines for data of new instruments, new \stan\ models with different kinds of regularisation, and/or addition of new template libraries. The code and documentation can be downloaded from the repository indicated in \S\ref{sec:implementation}.\looseness-2

The complexity in the kinematics observed in IFU surveys \citep[e.g.][]{krajnovic11}, but also numerical simulations \citep[e.g.][]{martig14,schulze17,walo20}, clearly indicates that a non-parametric approach is necessary to capture the great level of detail that current data offers. This has been shown for decades in early-type galaxies, but the potential is much greater in late-type spiral systems which display much more complex structures. In this respect, recent and upcoming large-scale IFU facilities (e.g. VIRUS-W, \citealt{fabricius08}; MEGARA, \citealt{megara} and WEAVE-LIFU, \citealt{weave}) operating at spectral resolutions above $R\ge5\,000$ open the door to explore the details of the LOSVDs in low velocity dispersion regimes (e.g. galaxy disks, dwarf galaxies), where it has been very hard to operate with current instrumentation. The non-parametric description of the LOSVDs will also have a big impact in the decomposition of galaxies into their kinematic/dynamical components. Current efforts rely heavily on the smooth LOSVDs provided by Gauss-Hermite parametrisations \citep[e.g.][]{tabor17,coccato18,oh20}. There is thus a great potential to go beyond those (necessary) efforts to explore galaxy mass assembly.\looseness-1

\begin{acknowledgements}
We thank the referee Prasenjit Saha for very useful comments that have helped improving the manuscript. We are grateful to Prashin Jethwa, Ignacio Mart\'in-Navarro, Marc Sarzi, Glenn van de Ven, and Eugene Vasiliev for many inspiring discussions over the course of this project. We are also indebted to Michela Rubino for her feedback and suggestions while testing earlier versions of the code, and Alireza Molaeinezhad for assisting in the preparation of the software package for Github. We thank Alessandro Pizzella for letting us use the IC\,0719 MUSE datacube in our analysis. J.~F-B thanks Andr\'es Asensio Ramos for introducing him to the world of bayesian inference methods and Stan in particular. We thank Michele Cappellari for letting us include some of his python functions in our software package. NGC\,4371 results are based on observations collected at the European Southern Observatory under ESO programme 060.A-9313(A).\newline
{\it Software acknowledgements}: Our code uses Astropy\footnote{\url{https://www.astropy.org/}} a community-developed core Python package for Astronomy \citep{astropy:2013,astropy:2018}, as well as NumPy\footnote{\url{https://numpy.org/}} \citep{oliphant06}, SciPy\footnote{\url{https://scipy.org/}} \citep{jones01} and Matplotlib\footnote{\url{https://matplotlib.org/}} \citep{hunter07}\newline
{\it Funding and financial support acknowledgements}: J.~F-B  acknowledges support through the RAVET project by the grant PID2019-107427GB-C32 from the Spanish Ministry of Science, Innovation and Universities (MCIU), and through the IAC project TRACES which is partially supported through the state budget and the regional budget of the Consejer\'ia de Econom\'ia, Industria, Comercio y Conocimiento of the Canary Islands Autonomous Community. 
\end{acknowledgements}
%

\bibliographystyle{aa}
\bibliography{bayes-losvd}

\appendix

\section{Trends with S/N ratio}
\label{app:losvd_types}

In this appendix we present the equivalent figures to Fig.~\ref{fig:losvd_types} for different S/N ratios. It is evident that the LOSVD recovery worsens as S/N decreases. It is interesting to see that not all the features are captured well even at S/N=100 (e.g. small bump on the positive wing of the case in the bottom row in all figures).

\begin{figure*}
   \centering
   \includegraphics[width=\linewidth]{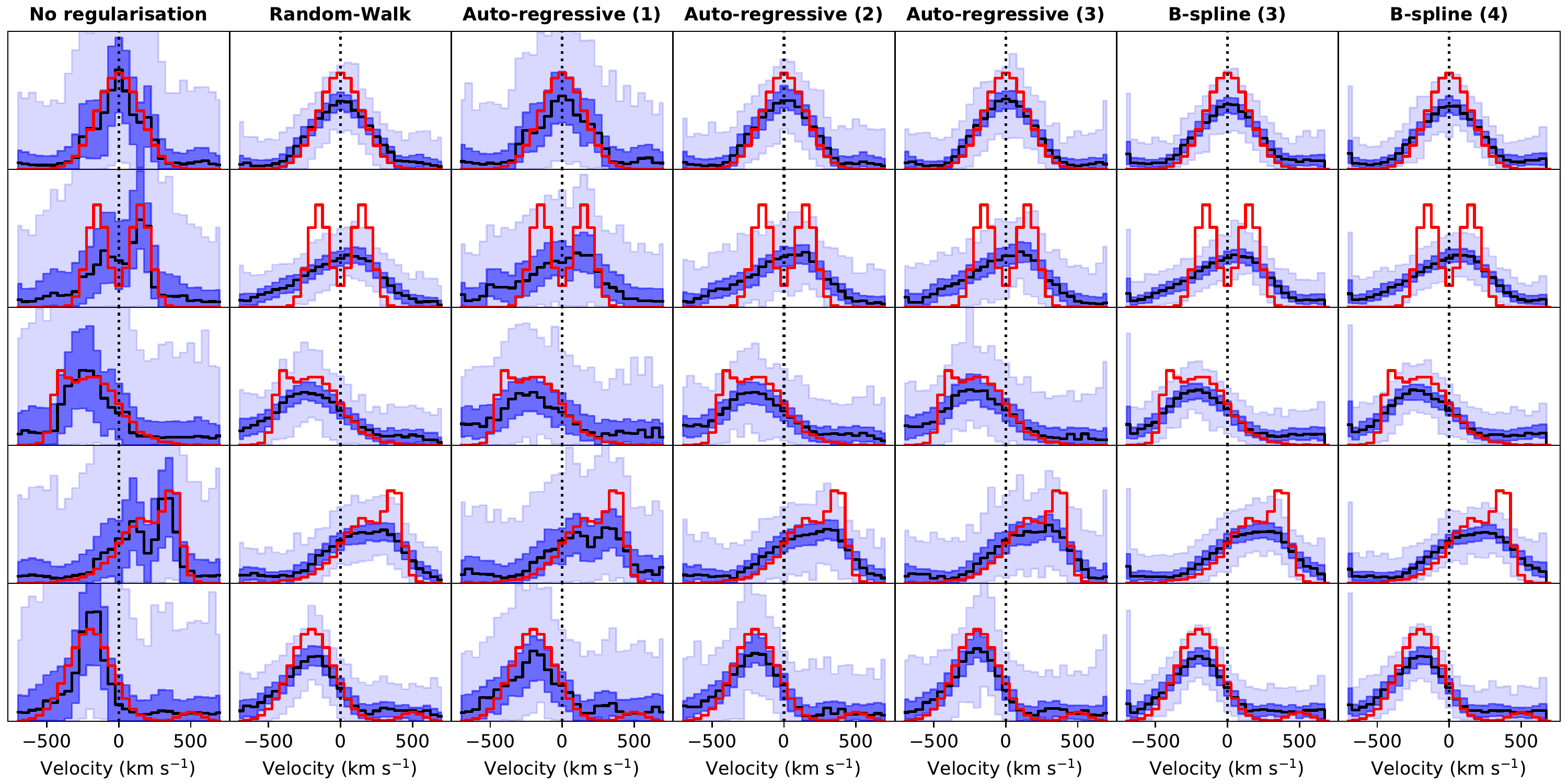}
   \caption{Same as Fig.~\ref{fig:losvd_types} for S/N=10.}
   \label{fig:losvd_types_10}
\end{figure*}

\begin{figure*}
   \centering
   \includegraphics[width=\linewidth]{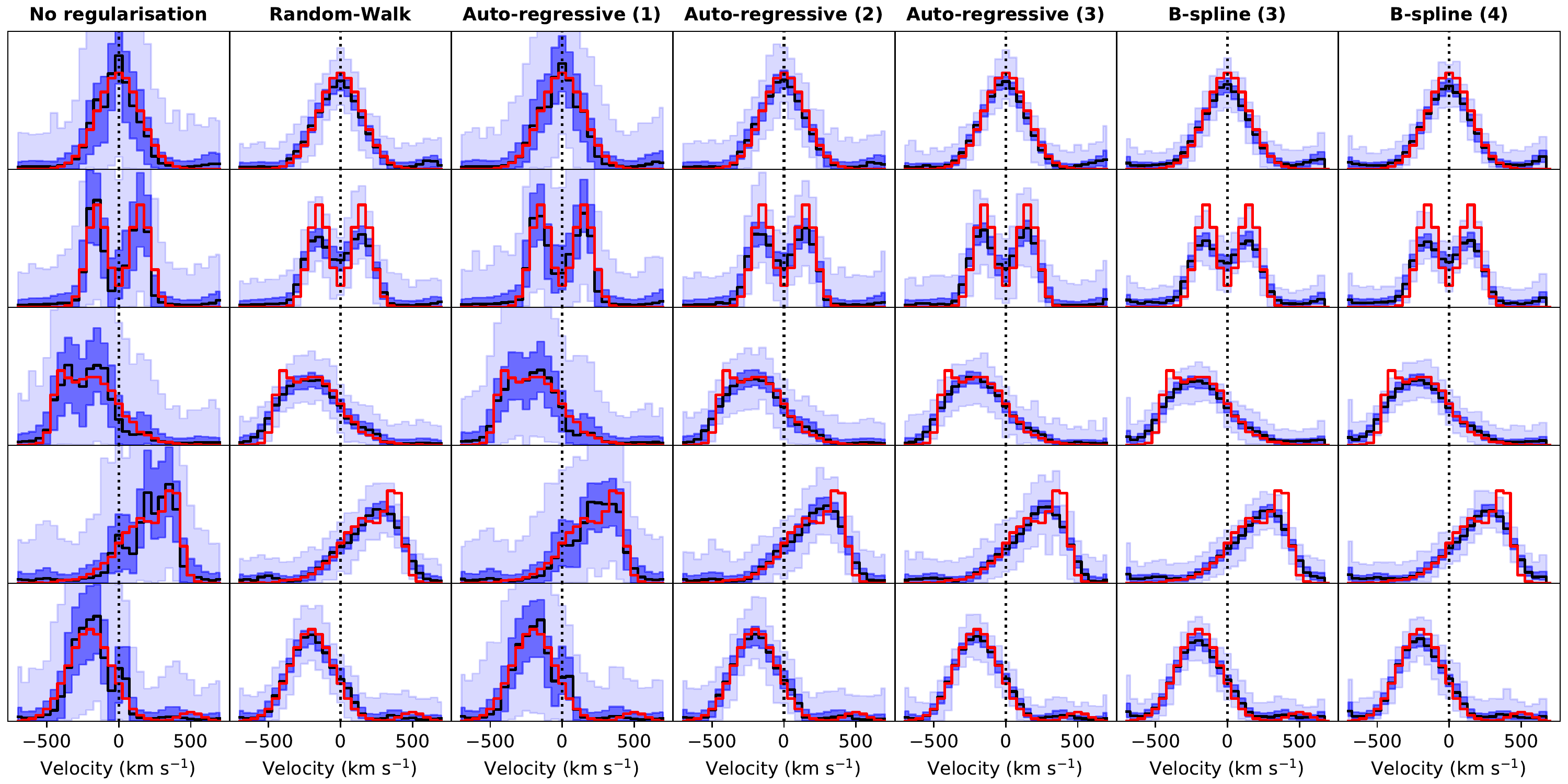}
   \caption{Same as Fig.~\ref{fig:losvd_types} for S/N=25.}
   \label{fig:losvd_types_25}
\end{figure*}

\begin{figure*}
   \centering
   \includegraphics[width=\linewidth]{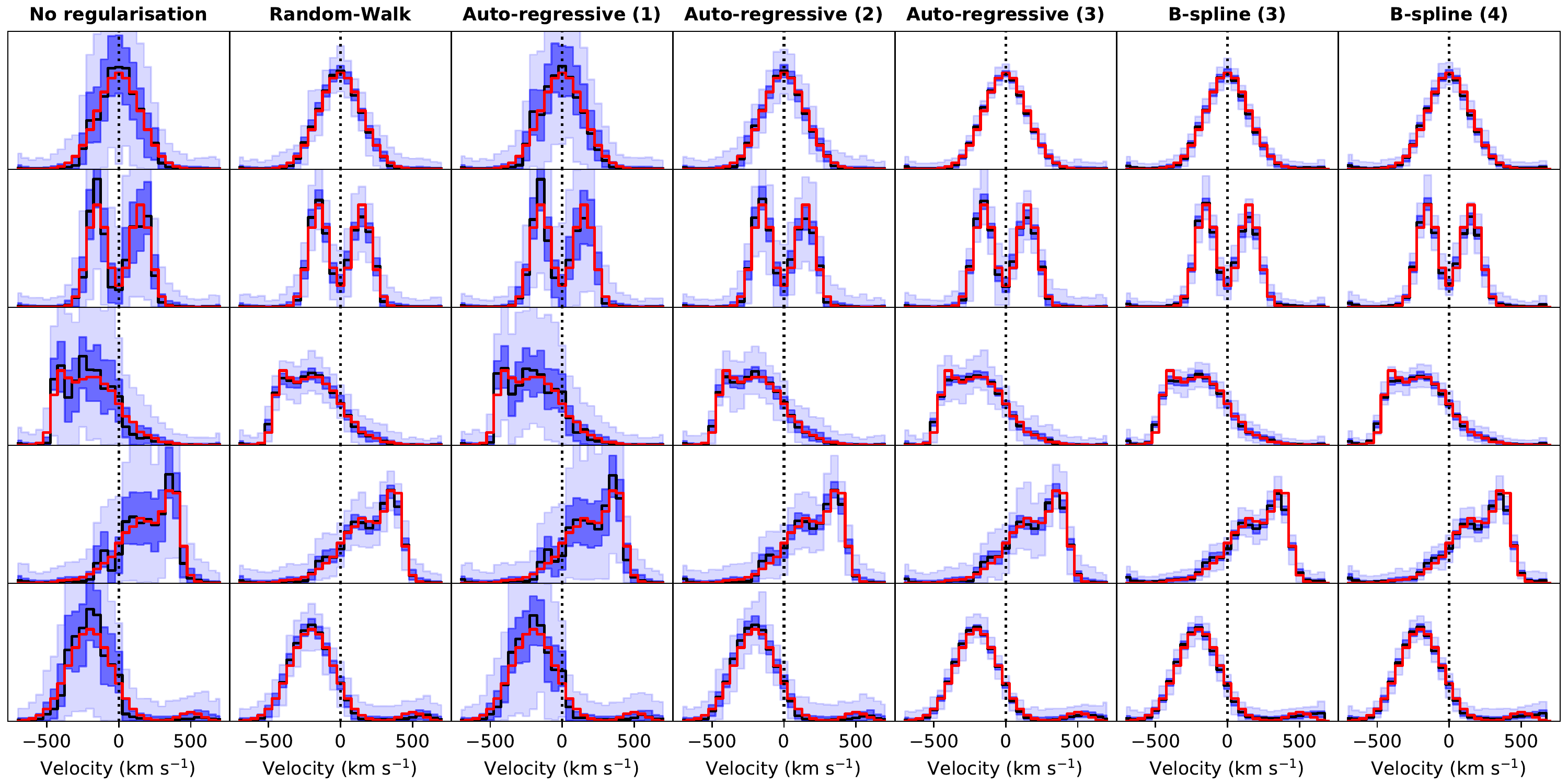}
   \caption{Same as Fig.~\ref{fig:losvd_types} for S/N=100.}
   \label{fig:losvd_types_50}
\end{figure*}

\begin{figure*}
   \centering
   \includegraphics[width=\linewidth]{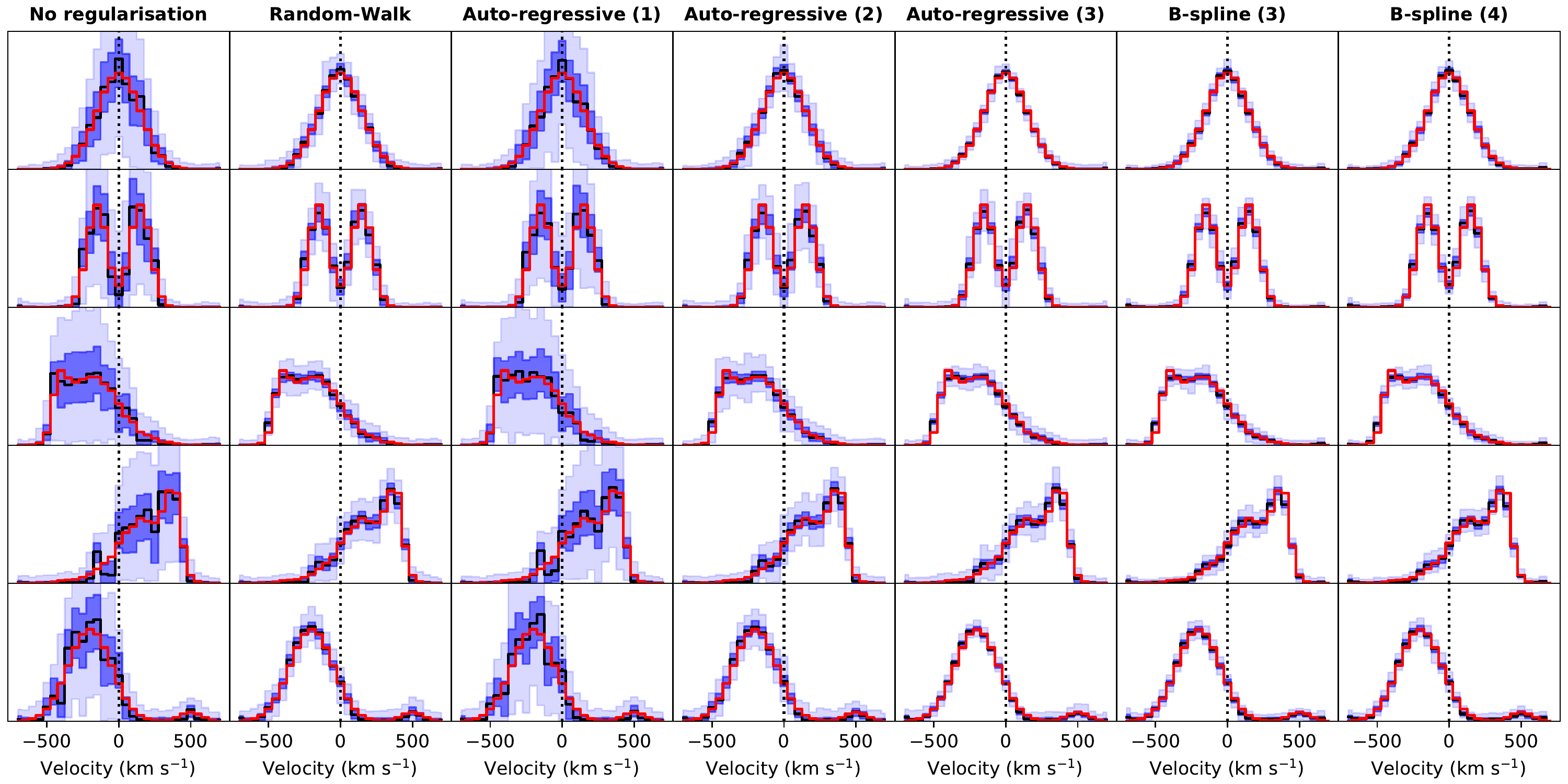}
   \caption{Same as Fig.~\ref{fig:losvd_types} for S/N=200.}
   \label{fig:losvd_types_100}
\end{figure*}

\end{document}